\documentclass{article}
\pdfoutput=1

\usepackage{PRIMEarxiv}

\usepackage[utf8]{inputenc} 
\usepackage[T1]{fontenc}    
\usepackage{hyperref}       
\usepackage{url}            
\usepackage{booktabs}       
\usepackage{amsfonts}       
\usepackage{nicefrac}       
\usepackage{algorithm} 
\usepackage{algpseudocode} %
\usepackage{microtype}      
\usepackage{lipsum}
\usepackage{fancyhdr}       
\usepackage{graphicx}       
\graphicspath{{media/}}     
\usepackage{tabularx}
\usepackage{multirow}
\usepackage{makecell}
\usepackage{amsmath}
\usepackage{graphicx}
\usepackage[table,xcdraw]{xcolor}

\title{MoLink: Distributed and Efficient Serving Framework for Large Models}

\author{
  Lewei Jin, Yongqi Chen, Kui Zhang, Yifan Zhuo, Yi Gao, Bowei Yang, Zhengong Cai, Wei Dong \\
   Zhejiang University \\
  \texttt{\{jinlewei,yqccchen,kuizhang,huoyf,gaoyi,boweiy,cstcaizg,dongw\}@zju.edu.cn} \\
}

\begin{document}
\maketitle

\begin{abstract}
Large language models represent a groundbreaking shift in generative AI. Yet, these advances come with a significant challenge: the high cost of model serving. To mitigate these costs, \textit{consumer-grade} GPUs emerge as a more affordable alternative. This presents an opportunity for more cost-efficient LLM serving by leveraging these GPUs.

However, it is non-trivial to achieve high-efficiency LLM serving on consumer-grade GPUs, mainly due to two challenges: 1) these GPUs are often deployed in limited network conditions; 2) these GPUs often exhibit heterogeneity in host systems.
To address these challenges, we present MoLink, a distributed LLM serving system for large models. It incorporates several key techniques, enabling efficient LLM serving on \textit{heterogeneous} and \textit{weakly connected} consumer-grade GPUs.
Our experiments demonstrate that it achieves throughput improvements of up to 458\% and cost-profit margin improvements of up to 151\%, compared to state-of-the-art systems.
MoLink allows users on Windows, Linux, and containerized VMs to seamlessly integrate GPUs with just a few lines of code over Ethernet or public networks. Currently, it supports 18 mainstream architectures of open-source large language models.
The source code is publicly available https://github.com/oldcpple/MoLink.

\end{abstract}

\section{Introduction}
Large language models represent a groundbreaking shift in generative AI, reshaping existing Internet services. Yet, these advances come with a significant challenge: the high cost of model serving.
For example, deploying the DeepSeek-671B model for inference requires 1.3TB of GPU memory. Using NVIDIA A100 GPUs (80GB memory each) \cite{a100}, at least 17 GPUs are needed just to meet basic deployment requirements. With each A100 priced at approximately 100,000 RMB, the hardware cost alone reaches 1.7 million RMB—a prohibitive expense for most small and medium-sized enterprises and research workstations.

To mitigate these costs, \textit{consumer-grade} GPUs emerge as a more affordable alternative. Taking the RTX 4090 \cite{4090} as an example, its computational performance (330 TFLOPS at FP16 precision) is comparable to that of the A100 (312 TFLOPS at FP16 precision), while providing one-fourth (24GB) of the A100's memory capacity (80GB) at only \textit{one-tenth} of its procurement cost. It is reported that the PC and AIB GPU market shipped over 101 million units in Q4 2021 alone \cite{gpus}. Given the vast availability of consumer-grade GPUs in the market, there is an opportunity for more cost-efficient LLM serving by leveraging these GPUs. 

However, achieving high-efficiency LLM serving on consumer-grade GPUs is non-trivial. First, \textit{these GPUs are typically distributed across geographic regions and connected via constrained networks}. Pipeline parallelism is often applied to overcome this network bottleneck, enabling distributed devices to handle specific model computation stages with low communication overhead. However, it suffers from pipeline bubbles due to 1) high micro-batch transmission delays and 2) transmission contention between prefill and decoding requests.


Second, \textit{these GPUs are typically deployed on diverse host systems}, ranging from Windows PCs and Linux servers to containerized VMs offered by cloud service providers \cite{AutoDL}. However, current cluster management systems (such as Kubernetes \cite{kubernetes}) are primarily designed for Linux environments and lack support for these heterogeneous systems. This limitation makes it difficult to schedule and optimize GPU resources automatically and efficiently, resulting in low utilization.

To address these challenges, we developed MoLink, a distributed LLM serving system for large models. It incorporates several key techniques such as dynamic micro-batch scheduling an chunk transmission, enableing efficient LLM serving on \textit{weak-connected} consumer-grade GPUs. 
Our experiment demonstrate that it achieves throughput
 improvements of up to 458\% and cost-profit margin improvements of up to 151\%, compared to state-of-the-art systems. 
 
MoLink provides flexible compute resource integration, allowing users in various environments (such as Linux, Windows, and AutoDL instances) to seamlessly integrate consumer-grade GPUs with just a few lines of code over Ethernet or public networks. Currently, MoLink supports mainstream open-source large language models (LLMs), including DeepSeek , Qwen, and Llama, delivering a low-cost, high-performance, and zero-barrier solution for deploying large model services. 

\section{Overview}

\begin{figure}[t]
    \centering
    \includegraphics[width=.95\linewidth]{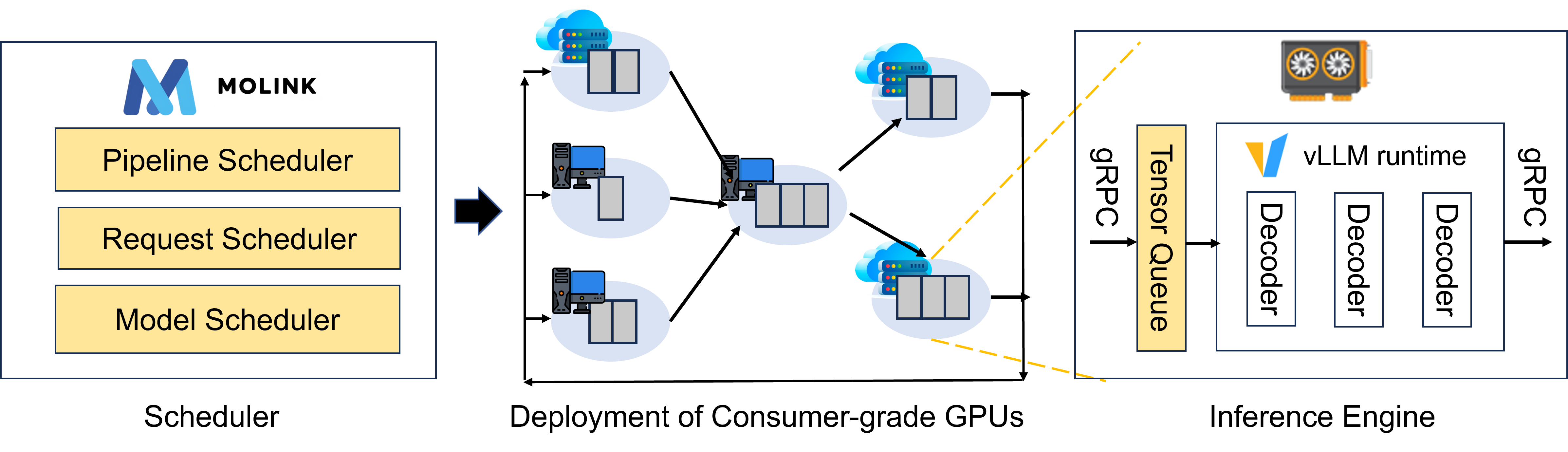}
    \caption{Overview of MoLink.}
\label{fig:overview}
\vspace{-2mm}
\end{figure}

MoLink is a distributed large model serving platform designed to harness consumer-grade computing devices. Figure \ref{fig:overview} shows the overview of MoLink. Compared to existing systems, MoLink offers three key advantages:

\textbf{(1) Distributed inference on consumer-grade GPUs}: MoLink facilitates efficient large-scale model inference through automated model partitioning and deployment. Its gRPC-based communication architecture supports cross-device deployment via Ethernet or public networks.

\textbf{(2) Flexible compute resource integration}: MoLink features a Kubernetes- and WSL-based distributed computing framework that enables seamless integration of heterogeneous devices, including Linux/Windows servers and containerized VMs (e.g., AutoDL instances \cite{AutoDL}).

\textbf{(3) High-performance serving}: MoLink employs advanced techniques such as dynamic micro-batch scheduling and prefill chunk transmission, delivering high concurrency performance even under constrained network conditions (e.g., 100Mbps bandwidth with high latency).

\subsection{Usage}
\label{sec:usage}
MoLink provides APIs for both model deployment and node management, the main APIs are listed as follow:

\subsubsection{Service Access API.}

\texttt{DeployLLMService(service\_name, model\_name,resource\_specification, inference\_parameters)}:
Deploys an LLM service with specified parameters: service\_name, model\_name, resource\_specification(e.g., GPU configuration), inference\_parameters. Deployment requests are forwarded to the Task Scheduler for node selection and model partitioning.

\texttt{GetAPIKey(service\_name) →  api\_key}:
Retrieves the authentication credential for an active LLM service.

\texttt{CheckServiceStatus (service\_name) →  status\_report}: Returns operational telemetry including service metadata (e.g. current state, uptime, request counts), aggregated performance metrics (e.g. throughput statistics, hardware utilization)

\texttt{DeleteLLMService (service\_name)}: Terminates the specified service and releases allocated hardware resources.

\subsubsection{Device Access API.}
\texttt{NodeAccess (node\_info)}: Registers a worker node into the cluster. Input requires a structured node descriptor including: hostname, OS type and version, hardware specification and network configuration.
We provide automation scripts to extract above local node attributes and invoke this API. 

\texttt{CheckNodeStatus (node\_name) → status\_report}: Retrieves metrics data from specified node, including: Metadata(e.g. IP address, hardware profile), resource utilization(e.g. GPU/CPU load, power consumption, temperatures).

\texttt{NodeExit (node\_name)}: Initiates graceful decommissioning of target node with the following steps: (1)Terminates active workloads, (2)Releases allocated resources, (3)Removes node from cluster registry.

\subsubsection{Supported Models} 
MoLink currently supports variant mainstream model architectures, listed in \ref{tab:models}.

\begin{table}[h]
\begin{tabular}{|l|l|l|l|}
\hline
BaichuanForCausalLM & BaichuanForCausalLM   & ChatGLMForCausalLM    & CohereForCausalLM \\ \hline
DeepseekForCausalLM & DeepseekV2ForCausalLM & DeepseekV3ForCausalLM & FalconForCausalLM \\ \hline
GemmaForCausalLM    & Gemma2ForCausalLM     & GlmForCausalLM        & GPT2LMHeadModel   \\ \hline
Phi3ForCausalLM     & QwenLMHeadModel       & Qwen2MoeForCausalLM   & Qwen2ForCausalLM  \\ \hline
Qwen3MoeForCausalLM & Qwen3ForCausalLM      &                       &                   \\ \hline
\end{tabular}
\caption{Supported models of MoLink.}
\label{tab:models}
\end{table}

\section{Design}

The MoLink framework employs a dual-node architecture comprising a master node and worker nodes.

\subsection{Master Node}
\begin{figure}[h]
    \centering
    \includegraphics[width=.5\linewidth]{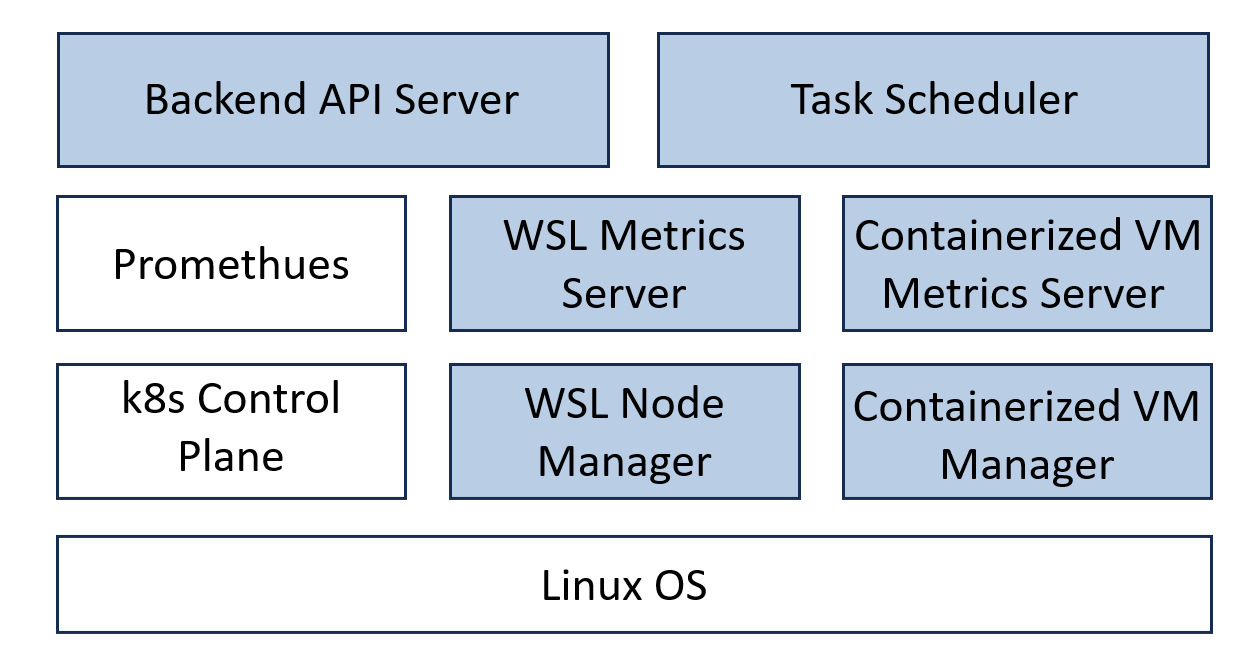}
    \caption{The design of master node.}
\label{fig:master_node}
\vspace{-2mm}
\end{figure}

Figure \ref{fig:master_node} shows the design of master node. It orchestrates cluster management, metrics data collection, and deployment task scheduling. Hosting on a Linux server, it integrates the following core components:

\subsubsection{Control Plane}
MoLink utilizes platform-specific components for node integration. The Kubernetes (k8s) Control Plane manages Linux node lifecycle operations (join/eviction) and schedules execution engine pods to target nodes based on directives from the Task Scheduler. To support Windows nodes and containerized VMs, we implement the WSL Node Manager and Containerized VM Manager, respectively. These components provide analogous functionality to the K8s Control Plane, handling node integration for their respective platforms and propagating deployment commands to target node daemons.

\textbf{Metrics Server}
MoLink employs Prometheus for periodic metrics data collection and storage from Linux nodes. Complementary components—the WSL Metrics Server and Containerized VM Metrics Server—deliver equivalent functionality for Windows nodes and containerized VMs, respectively.

\subsubsection{Task Scheduler}
Responsible for orchestrating deployment tasks for large language models (LLMs), primarily encompassing two core functions:

\textbf{(1) Node Selection:} When users submit deployment requests via the DeployLLMService API, they are required to specify the hardware requirements (GPU type and quantity). The Task Scheduler subsequently identifies nodes within the cluster that satisfy these constraints. It first determines the optimal parallelization strategy: preferentially selecting multi-GPU nodes compatible with Tensor Parallelism when available. When model deployment necessitates Pipeline Parallelism across multiple nodes, the scheduler prioritizes nodes exhibiting optimal inter-node latency and link bandwidth. 

\textbf{(2) Model Partitioning:} Upon node selection, the Task Scheduler determines the partitioning of model layers across the designated nodes. If all selected nodes possess identical GPU type and quantity, model layers will be partitioned uniformly. In heterogeneous node environments (e.g., nodes capable of multi-GPU TP alongside nodes with only a single GPU, or nodes with differing GPU types), a non-uniform partitioning strategy is employed. This strategy leverages real-time profiled performance metrics to preferentially allocate more layers to nodes with higher computational capacity. Furthermore, since the head node in the pipeline needs to launch the model inference API, and its execution engine is responsible for request scheduling and pipeline control, nodes with superior CPU performance and network connectivity are prioritized for this role. Following the determination of node selection and model partitioning strategies, the Task Scheduler dispatches the scheduling request to node management components (e.g., the K8ss Control Plane or WSL Node Manager). These components subsequently issue commands to the respective nodes to launch the engines during the execution phase.

Besides this, MoLink employs a Backend API Server to provide node management and model deployment APIs, as described in Section \ref{sec:usage}. 

\subsection{Worker Node}

\begin{figure}[t]
    \centering
    \includegraphics[width=1\linewidth]{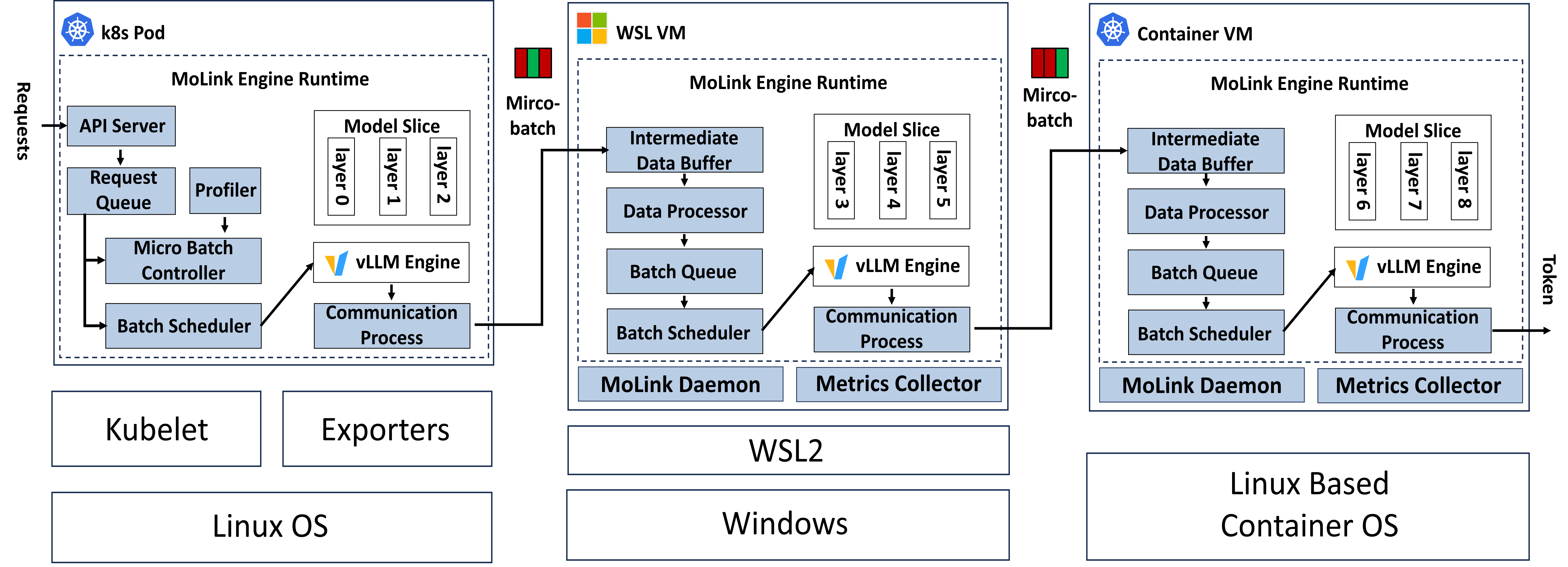}
    \caption{The design of worker node.}
\label{fig:worker_node}
\vspace{-2mm}
\end{figure}

Figure \ref{fig:worker_node} shows the design of worker node. It hosts distributed inference engines and execute model inference. Their implementation is detailed below across two dimensions: node integration and inference engine.

\subsubsection{Node Integration}
MoLink supports three node types: Linux, Windows, and Containerized VMs.
 
\textbf{(1) Linux Nodes:} Typically instantiated as physical Linux servers or cloud ECS instances. Node management is delegated to Kubelet, which:
(1) Receives scheduling directives from the master node's K8s Control Plane,
(2) Launches inference engines within pods using container images, injecting deployment configurations (e.g., assigned model layers, inference parameters), and
(3) Periodically synchronizes node state with the control plane.
For hardware monitoring, Linux nodes deploy exporter components (e.g., Node Exporter for CPU/network I/O,DCGM for GPUs). The master node's Prometheus periodically scrapes metrics from these exporters.

\textbf{(2) Windows Nodes:} To circumvent Kubernetes limitations and vLLM's Linux dependency on Windows hosts, we leverage WSL2-based Ubuntu 22.04 virtual machines. We provide automated scripts for integrating these VMs to the cluster. Within each WSL VM:
The MoLink Daemon (counterpart to Linux's Kubelet) receives commands from the master's WSL Node Manager, directly launching inference engines in the local environment (bypassing containerized pod launches).
The Metrics Collector (functionally analogous to Linux exporters) gathers CPU, network I/O, and GPU metrics via psutil, GPUtil, and pyCUDA, reporting to the master's WSL Metrics Server.

\textbf{(3) Containerized VMs:} Represent cost-effective instances from providers (e.g., autoDL, vast.ai) with constrained network topologies. These instances:
(1) Operate exclusively as isolated containers (e.g., k8s pods) with virtual IPs/ports,
(2) Permit only outbound internet access (e.g., for SSH connections), blocking inbound initiation by external nodes.
To enable cross-instance communication, our Daemon implements SSH tunnel port forwarding between nodes, repurposing the open SSH port for data transfer. The Daemon and Metrics Collector mirror their WSL counterparts in functionality and implementation.

\subsubsection{Distributed Inference Engine}
Inference engines are launched on the computing nodes,  differentiated to head engine and others. The head engine, acts as the service entry point and pipeline control plane, hosts the inference API Server  receives and parss user query requests. After preprocessing, user requests are pushed into the Request Queue. The Profiler profiles the computation latencies during engine startup, and network latency between nodes periodically. The Micro Batch Controller decides an ideal number of micro batches and the batch size(or number of batched tokens) at the start of each iteration, to better optimize the GPU bubbles, which is described in detailed in Section \ref{sec:dynamic_microbatch}. The Batch Scheduler is implemented on top of vLLM's continuous batching, which specifically schedules requests into micro batches. 

On a lower level, the LLM weights are hosted by the vLLM engine runtime, which takes micro batches as input and executes the inference computation. The output of the engine, namely the intermediate results, is then handled by the Communication Process, which conducts adaptive chunk transmission for intermediate results using gRPC. 

In subsequent engines, the intermediate results from last node are delivered into the Intermediate Data Buffer, deserialized by the Data Processor and then pushed into the Batch Queue. The Batch Scheduler schedules the batches in the order of arrival to the vLLM engine runtime, and the rest procedures are the same as in the head engine.

\section{Key Techniques}
\subsection{Dynamic Micro-batch scheduling}
\label{sec:dynamic_microbatch}
\begin{figure}[t]
    \centering
    \includegraphics[width=1\linewidth]{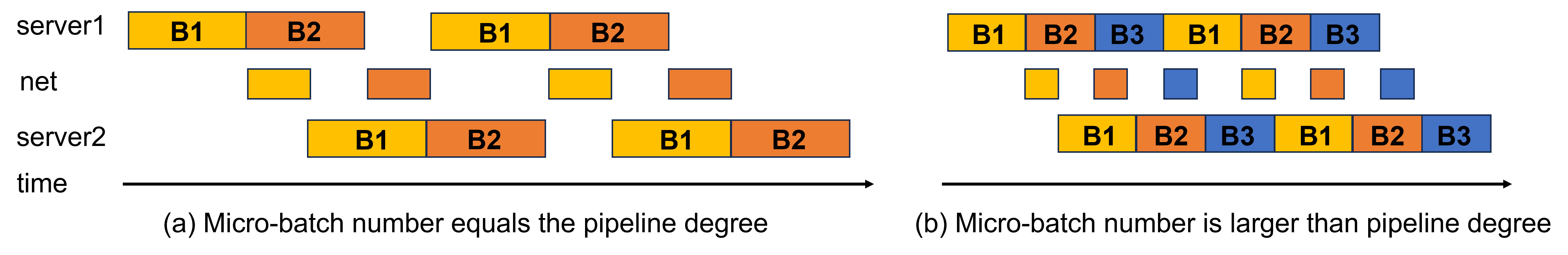}
    \vspace{-6mm}
    \caption{(a) Pipeline Bubble. (b) Dynamic Micro-batch scheduling.}
\label{fig:pipeline_bubble}
\vspace{-2mm}
\end{figure}

\textbf{Problem: Pipeline Bubble due to network latency.} To fully utilize GPUs, existing systems \cite{kwon2023efficient, yu2022orca} often set the number of micro-batches equal to the degree of pipeline parallelism. This approach assumes that the transmission overhead between servers is negligible, given the relatively small volume of activations. However, when servers are connected via limited bandwidth, the transmission overhead for activations can increase significantly. This can lead to pipeline bubbles, reducing the overall efficiency of the system.
 
Figure \ref{fig:pipeline_bubble}a shows a example when we use a micro-batch number equals to the pipeline degree. We can see that the transmission process of intermediate activations takes times and delays the execution of micro-batches (i.e., B1 and B2) in the target server. As a result, the server idles when finishing one of the micro-batch (i.e., B2) since the arrival of another micro-batch (i.e., B1) is always delayed. This under-utilization of servers naturally exists when the number of micro-batch is equals to the pipeline degree, since there is no more micro-batch fitting in the idle time of servers.  

\textbf{Solution: Dynamic Micro-batch scheduling.} 
To address the issue, We adjust the number of micro-batch to fill the pipeline bubbles. Figure \ref{fig:pipeline_bubble}b shows a example. By adding a new micro-batch (i.e., B3) that sequentially executing after the B2, we can see that the transmission time of B1 are overlapped by the execution of B3. Therefore reduce the idle time of servers. 

The optimal number of micro-batch can be different according to both the workload and network conditions, therefore, we dynamically adjust the number of micro-batches. 
The scheduling mechanism employs a two-phase profiling strategy: 
(1) Startup Profiling: Each engine measures computation latency across batch sizes and sequence lengths.
(2) Runtime Monitoring: Engines periodically profile downstream network conditions (latency, available bandwidth).

The head engine's Micro Batch Controller synthesizes these profiles to determine per-iteration parameters: (1) Constrains maximum batch size/token count for workload balance, (2) Computes an ideal micro-batch count $N$ through incremental search (starting at $N=1$) and (3) Terminates when the computation overhead of additional micro-batches can be fully overlapped with the residual bubbles. 

\subsection{Chunk Transmisssion}
\begin{figure}[t]
    \centering
    \includegraphics[width=.9\linewidth]{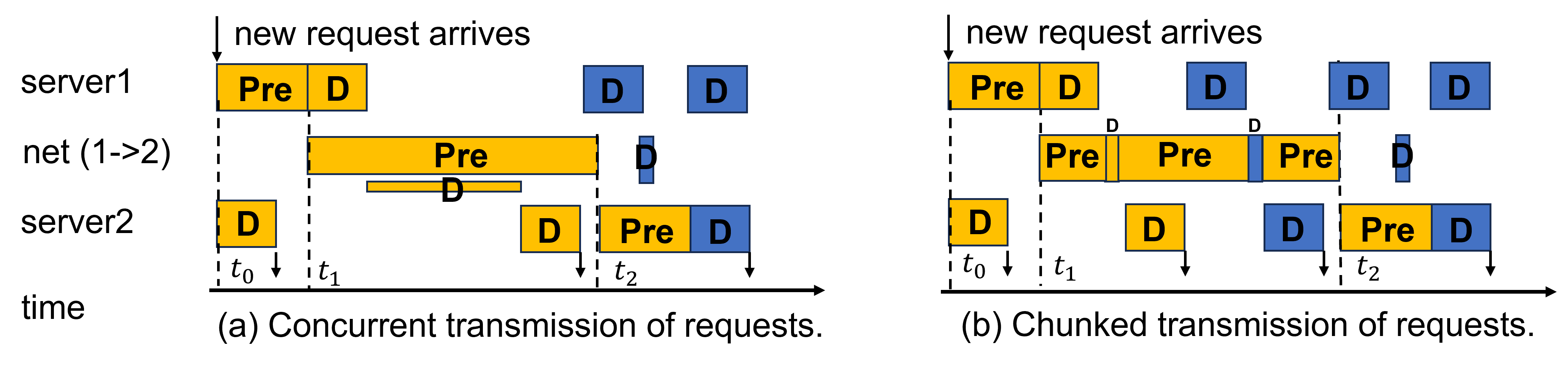}
    \caption{A case of transmission competition between prefill and decode for LLaMa-30B running on RTX 4090(s) linked with 100 mbps bandwidth. The prompt length is 1000. The batch size is 4.}
\label{fig:chunk_tranmission}
\vspace{-2mm}
\end{figure}
\textbf{Problem: Transmission competition between prefill and decode.} 
The competition often happens when sequentially processing a new arriving request in prefill phase and an existing batch of requests in decode phase. This is primarily due to the imbalance in transferred data volume between the two phases of inference: \textit{prefill} and \textit{decode}. 
Prefill requests can involve transmitting up to 1000 times more data than decode requests, as their volume scales with the number of processed tokens. 

Figure \ref{fig:chunk_tranmission}a shows an example. We can see that although the volume of decode is sent once the computation is finished at \( t_2 \), the transmission of decode is delayed by several milliseconds since a large part of the bandwidth is allocated to prefill. 
In real practice, we observe that the delay of decode can be hundreds of times greater compared to the transmission without competition.

\textbf{Solution: Chunk Transmisssion.} our idea is to transfer the  activation generated in prefill phase in chunks, which we referred as \textit{chunking transmission}.
Figure \ref{fig:chunk_tranmission}(b) shows an option of chunking and transmitting the activation of prefill requests. We can see that when prefill request finishes its execution, it start to transmit only a chunk of the activation generated. While this transmission is always finished before the transmission of decode, the decode are not delayed during its transmission. As a results, server2 finishes more iterations. 
For details, we continuously check whether there are requests finished in the server, and put their volume to transfer in the queue. The volume is divided into two type of queue depending on the phase of the requests. We priories activation transmission for decode requests. Therefore, we choose to transmit a decode request even there exits prefill requests needing transmission.  We currently use a fixed chunk size for the volume of the prefill requests.




\section{Evaluation}
\textbf{Baseline}
We conduct extensive experimental evaluations over MoLink and state-of-the-art distributed serving systems. 
\begin{itemize}
    \item Petals \cite{borzunov2023distributed}. The framework introduces serving LLMs with geo-distributed devices with consumer-grade GPUs. 
    \item vLLM \cite{kwon2023efficient}. It is a representative LLM serving system widely used in both academia and industry. It mainly designed for LLM serving in data center. It uses Ray to serving the LLMs with distributed devices. 
\end{itemize}

\textbf{Models}
We evaluate MoLink on LLaMa, a representative and popular open-source Transformer model family. Specifically, we use Llama2-7B \cite{llama2-7b} and  Llama2-70B \cite{llama2-70b} to study the system performance on models. For LLaMa2-7B, we run model inference with half-precision (FP16).  For Llama2-70B, we use 4-bit quantization to avoid high memory requirement that results in large pipeline parallelism size. 

\begin{figure}[t] 
    \centering 
    \begin{minipage}[t]{0.30\textwidth} 
        \centering
        \includegraphics[width=\linewidth]{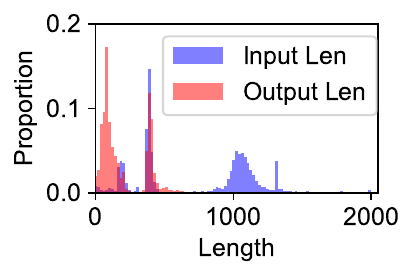} 
        \vspace{-8mm}
        \caption{Distribution of input and output length. } 
        \label{fig:distribution} 
    \end{minipage}
    \begin{minipage}[t]{0.30\textwidth} 
        \centering
        \includegraphics[width=\linewidth]{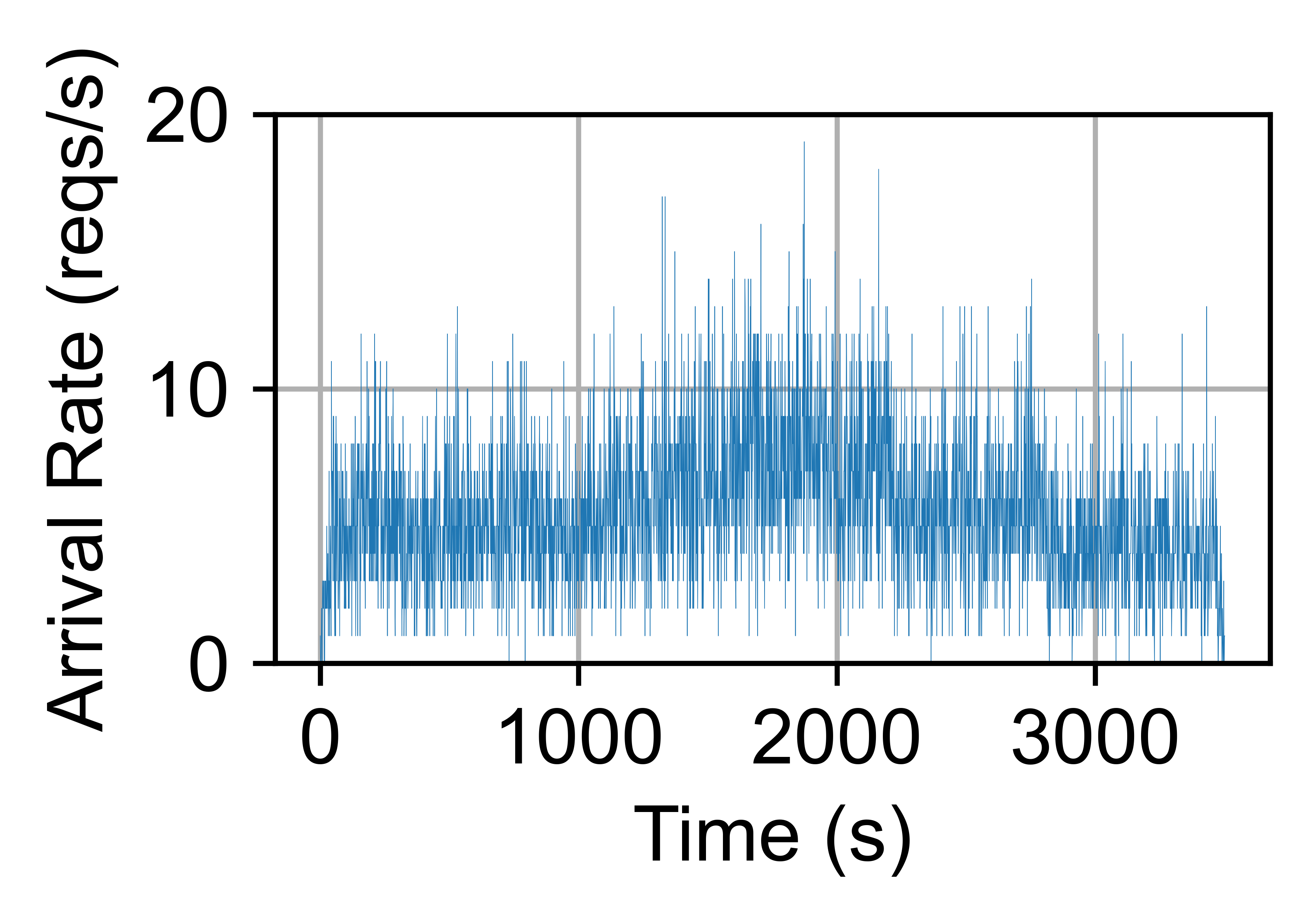} 
        \vspace{-8mm}
        \caption{Arrival rate} 
        \label{fig:arr_rate} 
    \end{minipage}
    \vspace{-2mm}
\end{figure}

\textbf{Traces.}
We use Azure Conversation \cite{azureconversation} to simulate the arrival of requests. 
Fig. \ref{fig:distribution} shows the length distribution of the datasets. 
Fig. \ref{fig:arr_rate} shows the arrival rate of the datasets. 
We remove requests with input lengths larger than 256 or output lengths larger than 512, and scale the frequency of requests arrival to fit in the GPUs we use, which we indicate as the arrival rate in our experiment. 

\textbf{Metrics}
We measure the performance of LLM service using three metrics: 
(1) Time to First Token (TTFT), which indicates the average time to first token, which typically including both the queue time and the duration of the prefill phase. 
(2) Time per Output Token (TPOT), which represents the average time taken to generate each token after the first one. 
(3) Throughput, which represents the average number of generated tokens every seconds.

\subsection{Overall Performance}
We evaluate our system on consumer-grade GPUs (e.g., NVIDIA RTX 4090) using two servers, each equipped with one RTX 4090 GPU. Pipeline parallelism is employed for distributed inference. To simulate real-world deployment conditions, we vary the interconnecting bandwidth and latency. 
The results are shown in Table \ref{tab:main_performance}. We can see that: (1) MoLink achieves a throughput up to 458\% higher than Petals and up to 44\% higher than vLLM. (2) MoLink reduces TTFT by 17.6\% on average compared to Petals and 8.4\% compared to vLLM. (3) MoLink achieves an average 22.8\% reduction over Petals and a 41.5\% reduction over vLLM.

\begin{table}[t]
\centering
\resizebox{\textwidth}{!}{
\begin{tabular}{cc|ccc|ccc|ccc}
\hline
                                                     &                                                         & \multicolumn{3}{c|}{\begin{tabular}[c]{@{}c@{}}Throughput \\ (tokens / s)\end{tabular}} & \multicolumn{3}{c|}{TTFT (s)}                      & \multicolumn{3}{c}{TPOT (s)}                       \\ \hline
\multicolumn{1}{c|}{Network}                         & \begin{tabular}[c]{@{}c@{}}trace\\ (req/s)\end{tabular} & \textbf{Petals}              & \textbf{vLLM}             & \textbf{MoLink}              & \textbf{Petals} & \textbf{vLLM}  & \textbf{MoLink} & \textbf{Petals} & \textbf{vLLM}  & \textbf{MoLink} \\ \hline
\multicolumn{1}{c|}{\multirow{2}{*}{100Mbps / 10ms}} & 0.5                                                     & 23.438                       & 89.68                     & \color[HTML]{3166FF}\textbf{106.238}             & 0.261           & 0.245          & \color[HTML]{3166FF}\textbf{0.231}  & 0.116           & 0.247          & \color[HTML]{3166FF}\textbf{0.102}  \\
\multicolumn{1}{c|}{}                                & 1                                                       & 45.343                       & 135.86                    & \color[HTML]{3166FF}\textbf{196.142}             & 0.279           & 0.264          & \color[HTML]{3166FF}\textbf{0.243}  & 0.290           & 0.348          & \color[HTML]{3166FF}\textbf{0.255}  \\ \hline
\multicolumn{1}{c|}{\multirow{2}{*}{1Gbps / 10ms}}   & 0.5                                                     & 26.444                       & 91.65                     & \color[HTML]{3166FF}\textbf{111.162}             & 0.204           & \color[HTML]{3166FF}\textbf{0.177} & 0.181           & 0.193           & \color[HTML]{3166FF}\textbf{0.168} & 0.170           \\
\multicolumn{1}{c|}{}                                & 1                                                       & 47.068                       & 163.95                    & \color[HTML]{3166FF}\textbf{207.973}             & 0.209           & \color[HTML]{3166FF}\textbf{0.180} & 0.189           & 0.162           & 0.198          & \color[HTML]{3166FF}\textbf{0.144}  \\ \hline
\multicolumn{1}{c|}{\multirow{2}{*}{10Gbps / 10ms}}  & 0.5                                                     & 21.332                       & 92.13                     & \color[HTML]{3166FF}\textbf{119.11}              & 0.200           & 0.182          & \color[HTML]{3166FF}\textbf{0.130}  & 0.097           & 0.176          & \color[HTML]{3166FF}\textbf{0.056}  \\
\multicolumn{1}{c|}{}                                & 1                                                       & 41.878                       & 215.41                    & \color[HTML]{3166FF}\textbf{221.56}              & 0.199           & 0.175          & \color[HTML]{3166FF}\textbf{0.148}  & 0.107           & 0.183          & \color[HTML]{3166FF}\textbf{0.056}  \\ \hline
\end{tabular}
}
\label{tab:main_performance}
\caption{The main performance results of MoLink against state-of-the art systems. The \textcolor[HTML]{3166FF}{blue} indicates the best performance. }
\end{table}

\subsection{Cost profit margin analysis}
We evaluate the economic efficiency of deploying consumer-grade GPUs (e.g., RTX 4090) versus professional GPUs (e.g., H20, A100) by analyzing the cost-profit margin, defined as: 

\[
\textit{Cost-Profit Margin} = (Profit - Cost) / Cost
\]

where the profit is from generating tokens from the serving systems, the cost is the cost to maintain the devices.  For LLMs inference serving, the profit can be calculated as:

\[
Profit = \textit{Throughtput} * \textit{Token price}
\]

We now consider two calculation of cost.

\textbf{(1) Purchase and maintain the devices locally.} In this scenario, we consider both the hardware procurement costs and the electricity consumption expenses. We assume a 5-year usage period, so the total cost is calculated as follows: 

\[
\textit{Cost = Price of devices / 5 years + Power price * Power Consumption}
\]

(2) Rent the cloud devices. In this situation, we consider only the rent price of devices provided by the cloud service platform , so the total cost is calculated as follows: 

\[
Cost = \textit{Price of cloud servers}
\]


\begin{table}[t]
\centering
\begin{tabular}{
>{\columncolor[HTML]{FFFFFF}}c 
>{\columncolor[HTML]{FFFFFF}}c 
>{\columncolor[HTML]{FFFFFF}}c |
>{\columncolor[HTML]{FFFFFF}}c 
>{\columncolor[HTML]{FFFFFF}}c 
>{\columncolor[HTML]{FFFFFF}}c |
>{\columncolor[HTML]{FFFFFF}}c 
>{\columncolor[HTML]{FFFFFF}}c 
>{\columncolor[HTML]{FFFFFF}}c }
\hline
\multicolumn{3}{c|}{\cellcolor[HTML]{FFFFFF}}                                                                                                                                                                                                                                                      & \multicolumn{3}{c|}{\cellcolor[HTML]{FFFFFF}System throughput (tokens / s)} & \multicolumn{3}{c}{\cellcolor[HTML]{FFFFFF}Cost profit margin}            \\ \hline
\multicolumn{1}{c|}{\cellcolor[HTML]{FFFFFF}Hardware}                                                                                               & \multicolumn{1}{c|}{\cellcolor[HTML]{FFFFFF}Network}                          & \begin{tabular}[c]{@{}c@{}}trace\\      (req/s)\end{tabular} & \textbf{Petals}  & \textbf{vLLM}  & \textbf{MoLink}                         & \textbf{Petals} & \textbf{vLLM} & \textbf{MoLink}                         \\ \hline
\multicolumn{1}{c|}{\cellcolor[HTML]{FFFFFF}}                                                                                                       & \multicolumn{1}{c|}{\cellcolor[HTML]{FFFFFF}}                                 & 0.5                                                          & 23.438           & 89.68          & {\color[HTML]{3166FF} \textbf{106.238}} & -80.3\%         & -24.5\%       & {\color[HTML]{3166FF} \textbf{-10.6\%}} \\
\multicolumn{1}{c|}{\cellcolor[HTML]{FFFFFF}}                                                                                                       & \multicolumn{1}{c|}{\multirow{-2}{*}{\cellcolor[HTML]{FFFFFF}100Mbps / 10ms}} & 1                                                            & 45.343           & 135.86         & {\color[HTML]{3166FF} \textbf{196.142}} & -61.8\%         & 14.3\%        & {\color[HTML]{3166FF} \textbf{65.1\%}}  \\ \cline{2-9} 
\multicolumn{1}{c|}{\cellcolor[HTML]{FFFFFF}}                                                                                                       & \multicolumn{1}{c|}{\cellcolor[HTML]{FFFFFF}}                                 & 0.5                                                          & 26.444           & 91.65          & {\color[HTML]{3166FF} \textbf{111.162}} & -77.7\%         & -22.9\%       & {\color[HTML]{3166FF} \textbf{-6.5\%}}  \\
\multicolumn{1}{c|}{\cellcolor[HTML]{FFFFFF}}                                                                                                       & \multicolumn{1}{c|}{\multirow{-2}{*}{\cellcolor[HTML]{FFFFFF}1Gbps / 10ms}}   & 1                                                            & 47.068           & 163.95         & {\color[HTML]{3166FF} \textbf{207.973}} & -60.4\%         & 38.0\%        & {\color[HTML]{3166FF} \textbf{75.0\%}}  \\ \cline{2-9} 
\multicolumn{1}{c|}{\cellcolor[HTML]{FFFFFF}}                                                                                                       & \multicolumn{1}{c|}{\cellcolor[HTML]{FFFFFF}}                                 & 0.5                                                          & 21.332           & 92.13          & {\color[HTML]{3166FF} \textbf{105.687}} & -82.0\%         & -22.5\%       & {\color[HTML]{3166FF} \textbf{-11.1\%}} \\
\multicolumn{1}{c|}{\multirow{-6}{*}{\cellcolor[HTML]{FFFFFF}\begin{tabular}[c]{@{}c@{}}Two Server,\\ each server has a \\ RTX  4090\end{tabular}}} & \multicolumn{1}{c|}{\multirow{-2}{*}{\cellcolor[HTML]{FFFFFF}10Gbps / 10ms}}  & 1                                                            & 41.878           & 215.41         & {\color[HTML]{3166FF} \textbf{221.56}}  & -64.8\%         & 81.3\%        & {\color[HTML]{3166FF} \textbf{86.5\%}}  \\ \hline
\end{tabular}
\label{tab:cost_profit_margin_weak}
\caption{The cost profit margin for locally maintaining the devices. The API price of LLama2-7B is \$0.25 \cite{apiprice} (1.8 yuan) / per million output tokens. The power consumption for RTX 4090 is 450W. The electricity price is 0.538 yuan / KWh. The price of RTX 4090 is 12999 yuan. The \textcolor[HTML]{3166FF}{blue} indicates the best performance. }
\end{table}

\begin{table}[t]
\centering
\resizebox{\textwidth}{!}{
\begin{tabular}{c|ccc|ccc|ccc|ccc}
\hline
{\color[HTML]{000000} }          & \multicolumn{3}{c|}{{\color[HTML]{000000} \begin{tabular}[c]{@{}c@{}}Throughput\\ (tokens / s)\end{tabular}}}            & \multicolumn{3}{c|}{{\color[HTML]{000000} TTFT (s)}}                                                                      & \multicolumn{3}{c|}{{\color[HTML]{000000} TPOT (s)}}                                                               & \multicolumn{3}{c}{{\color[HTML]{000000} Cost profit margin}}                                                 \\ \hline
{\color[HTML]{000000} trace}     & {\color[HTML]{000000} \textbf{S1}}     & {\color[HTML]{000000} \textbf{S2}}     & {\color[HTML]{000000} \textbf{S3}}     & {\color[HTML]{000000} \textbf{S1}}      & {\color[HTML]{000000} \textbf{S2}}     & {\color[HTML]{000000} \textbf{S3}}     & {\color[HTML]{000000} \textbf{S1}}    & {\color[HTML]{000000} \textbf{S2}}    & {\color[HTML]{000000} \textbf{S3}} & {\color[HTML]{000000} \textbf{S1}}    & {\color[HTML]{000000} \textbf{S2}} & {\color[HTML]{000000} \textbf{S3}}    \\ \hline
{\color[HTML]{000000} 1 req / s} & {\color[HTML]{3166FF} \textbf{193.39}} & {\color[HTML]{000000} 185.66}          & {\color[HTML]{000000} 149.61}          & {\color[HTML]{000000} 0.163}            & {\color[HTML]{000000} 0.165}           & {\color[HTML]{000000} 0.572}           & {\color[HTML]{000000} 0.046}          & {\color[HTML]{000000} 0.047}          & {\color[HTML]{000000} 0.362}       & {\color[HTML]{3166FF} \textbf{82\%}}  & {\color[HTML]{000000} 53\%}        & {\color[HTML]{000000} 31\%}           \\
{\color[HTML]{000000} 3 req / s} & {\color[HTML]{000000} 426.31}          & {\color[HTML]{3166FF} \textbf{461.79}} & {\color[HTML]{000000} 442.15}          & {\color[HTML]{000000} 0.874}            & {\color[HTML]{000000} 0.212}           & {\color[HTML]{000000} 0.558}           & {\color[HTML]{000000} 0.484}          & {\color[HTML]{000000} 0.071}          & {\color[HTML]{000000} 0.408}       & {\color[HTML]{3166FF} \textbf{302\%}} & {\color[HTML]{000000} 281\%}       & {\color[HTML]{000000} 287\%}          \\
{\color[HTML]{000000} 5 req / s} & {\color[HTML]{000000} 264.97}          & {\color[HTML]{000000} 606.48}          & {\color[HTML]{3166FF} \textbf{737.74}} & {\color[HTML]{FE0000} \textbf{26.457}}  & {\color[HTML]{FE0000} \textbf{1.469}}  & {\color[HTML]{000000} 0.53}            & {\color[HTML]{FE0000} \textbf{1.6}}   & {\color[HTML]{000000} 0.772}          & {\color[HTML]{000000} 0.438}       & {\color[HTML]{000000} 150\%}          & {\color[HTML]{000000} 400\%}       & {\color[HTML]{3166FF} \textbf{546\%}} \\
{\color[HTML]{000000} 6 req / s} & {\color[HTML]{000000} 279.28}          & {\color[HTML]{000000} 604.94}          & {\color[HTML]{3166FF} \textbf{854.78}} & {\color[HTML]{FE0000} \textbf{78.047}}  & {\color[HTML]{FE0000} \textbf{16.33}}  & {\color[HTML]{000000} 0.589}           & {\color[HTML]{FE0000} \textbf{1.776}} & {\color[HTML]{000000} 0.844}          & {\color[HTML]{000000} 0.526}       & {\color[HTML]{000000} 163\%}          & {\color[HTML]{000000} 399\%}       & {\color[HTML]{3166FF} \textbf{648\%}} \\
{\color[HTML]{000000} 7 req / s} & {\color[HTML]{000000} 429.6}           & {\color[HTML]{000000} 594.58}          & {\color[HTML]{3166FF} \textbf{927.36}} & {\color[HTML]{FE0000} \textbf{119.708}} & {\color[HTML]{FE0000} \textbf{29.976}} & {\color[HTML]{000000} 0.601}           & {\color[HTML]{FE0000} \textbf{1.983}} & {\color[HTML]{000000} 0.983}          & {\color[HTML]{000000} 0.583}       & {\color[HTML]{000000} 305\%}          & {\color[HTML]{000000} 391\%}       & {\color[HTML]{3166FF} \textbf{711\%}} \\
{\color[HTML]{000000} 8 req / s} & {\color[HTML]{000000} 405.1}           & {\color[HTML]{000000} 342.85}          & {\color[HTML]{3166FF} \textbf{909.16}} & {\color[HTML]{FE0000} \textbf{167.048}} & {\color[HTML]{FE0000} \textbf{44.678}} & {\color[HTML]{FE0000} \textbf{14.483}} & {\color[HTML]{FE0000} \textbf{2.181}} & {\color[HTML]{FE0000} \textbf{1.195}} & {\color[HTML]{000000} 0.745}       & {\color[HTML]{000000} 282\%}          & {\color[HTML]{000000} 183\%}       & {\color[HTML]{3166FF} \textbf{696\%}} \\ \hline
\end{tabular}
}
\label{tab:cost_profit_margin}
\caption{The cost profit margin for renting the cloud devices. The API price is \$2.75 \cite{apiprice} / per millison output tokens. The price of a RTX 4090 is \$0.26 / per hour. The price of a A100(80G) is \$1.16 / per hour.
The price of a H20(96G) is \$1.05 / per hour. They are rent from AutoDL. The \textcolor[HTML]{FE0000}{red} indicates the latency that exceeds 1s. The \textcolor[HTML]{3166FF}{blue} indicates the best performance. }
\end{table}

\textbf{Purchase and locally maintain the devices.} We compare Molink against Petals and vLLM under different network settings on Llama-2-7B on cost profit margin. The results are shown in Table \ref{tab:cost_profit_margin_weak}. 
We can see that MoLink achieves the highest cost-profit margin, outperforming Petals by up to 151\% and vLLM by 50.7\%.

\textbf{Rent the cloud devices.} We compare Molink against vLLM under different hardware settings on Llama-2-70B-AWQ(4bit). All the device is rent from cloud provider AutoDL \cite{AutoDL}. 
\begin{itemize}
    \item S1 (vLLM on H20-96GB) – Single-server deployment with an NVIDIA H20 (96GB). 
    \item S2 (vLLM on A100-80GB) – Single-server deployment with an NVIDIA A100 (80GB).
    \item S3 (MoLink on 4×RTX 4090-24GB) – Distributed deployment across four servers, each with an RTX 4090 (24GB), using pipeline parallelism.
\end{itemize}

The results are shown in Table \ref{tab:cost_profit_margin}. We can see that: (1) Distributed deployment across multiple consumer-grade GPUs (e.g., RTX 4090) demonstrates significantly higher workload capacity compared to single-node setups, achieving a peak throughput of 7 requests per second (req/s). (2) Consumer-grade GPUs (e.g., RTX 4090) achieve up to 75\% higher cost-profit margins compared to data-center-grade GPUs (e.g., NVIDIA A100).

\section{Related work}
\textbf{Machine Learning Model Serving.} Many recent LLM-specific systems tackle the unpredictable execution time and high memory consumption in LLM serving. Orca \cite{yu2022orca} proposed iteration level scheduling to release resources once a request is finished. Speculative Inference \cite{leviathan2023fast,miao2023specinfer} applies a small model to predict multiple output tokens. Splitwise \cite{patel2024splitwise} and DistServe \cite{zhong2024distserve} disaggregates the prompt and decode phase of requests. Sarathi \cite{agrawal2023sarathi} introduce chunked prefill. The key distinction is that Sarathi-Serve splits prefill execution into multiple tasks (where each chunk corresponds to a task), whereas MoLink splits the transmission of prefill results into chunks. All the above works are orthogonal to our work. 

\textbf{Distributed LLM Serving.} 
Several works have explored the potential of utilizing distributed GPUs for ML tasks. Some approaches \cite{jia2022whale,park2020hetpipe} co-design model partitioning and placement strategies for heterogeneous clusters. Others like Learninghome \cite{ryabinin2020towards} and DeDLOC \cite{diskin2021distributed} investigate network-aware routing in decentralized environments. SWARM \cite{ryabinin2023swarm} optimizes pipeline communication in heterogeneous networks, while additional efforts employ approximations to reduce either network communication \cite{wang2023cocktailsgd} or synchronization overhead \cite{hsieh2017gaia}. 
GPU selection strategies are addressed by SkyPilot \cite{yang2023skypilot} and Mélange \cite{griggs2024m}.  These works primarily focus on addressing device heterogeneity through model placement and scheduling, making them orthogonal to our approach.

\section{Conclusion}
In this paper, we present MoLink, a distributed LLM serving system for large models. It incorporates several key techniques, enabling efficient LLM serving on \textit{heterogeneous} and \textit{weakly connected} consumer-grade GPUs.
Our experiments demonstrate that it achieves throughput improvements of up to 458\% and cost-profit margin improvements of up to 151\%, compared to state-of-the-art systems.
MoLink allows users on Windows, Linux, and containerized VMs to seamlessly integrate GPUs with just a few lines of code over Ethernet or public networks. Currently, it supports 18 mainstream architectures of open-source large language models.

\bibliographystyle{unsrt}  
\bibliography{main}

\end{document}